\documentclass{article}

\usepackage[preprint,nonatbib]{nips_2018}

\usepackage[utf8]{inputenc} 
\usepackage[T1]{fontenc}    
\usepackage{hyperref}       
\usepackage{url}            
\usepackage{booktabs}       
\usepackage{amsfonts}       
\usepackage{nicefrac}       
\usepackage{microtype}      
\usepackage{graphicx}
\usepackage{authblk}
\usepackage{amsmath}

\usepackage{color}
\usepackage{kotex}
\title{Uncertainty quantification of molecular property prediction using Bayesian neural network models}

\author[1]{\textbf{Seongok Ryu}}
\author[2]{\textbf{Yongchan Kwon}}
\author[1,3]{\textbf{Woo Youn Kim}}
\affil[1]{Department of Chemistry, KAIST}
\affil[2]{Department of Statistics, Seoul National University}
\affil[3]{KAIST Institute for Artificial Intelligence}

\begin{document}

\maketitle

\begin{abstract}
In chemistry, deep neural network models have been increasingly utilized in a variety of applications such as molecular property predictions, novel molecule designs, and planning chemical reactions.
Despite the rapid increase in the use of state-of-the-art models and algorithms, deep neural network models often produce poor predictions in real applications because model performance is highly dependent on the quality of training data. 
In the field of molecular analysis, data are mostly obtained from either complicated chemical experiments or approximate mathematical equations, and then quality of data may be questioned.
In this paper, we quantify uncertainties of prediction using Bayesian neural networks in molecular property predictions.
We estimate both model-driven and data-driven uncertainties, demonstrating the usefulness of uncertainty quantification as both a quality checker and a confidence indicator with the three experiments.
Our results manifest that uncertainty quantification is necessary for more reliable molecular applications and Bayesian neural network models can be a practical approach.
\end{abstract}

\section{Introduction}

Modern deep neural network models have been used in various molecular applications, such as high-throughput screening for drug discovery \cite{gomes2017atomic, jimenez2018k, mayr2016deeptox, ozturk2018deepdta}, \textit{de novo} molecular design \cite{de2018molgan, gomez2018automatic, guimaraes2017objective, jin2018junction, kusner2017grammar, li2018learning, segler2017generating, you2018graph} and planning chemical reactions \cite{segler2018planning, wei2016neural, zhou2017optimizing}. Deep learning models show comparable and sometimes better performance than principle-based approaches in predicting molecular properties \cite{faber2017prediction, gilmer2017neural, schutt2017schnet, schutt2017quantum, smith2017ani}. 


A reliable statistical analysis critically depends on both the quality and the amount of data, however, most real datasets are often obtained from complicated chemical experiments in the field of chemistry, lacking both qualities and quantities.
For example, \cite{feinberg2018spatial} mentioned that more qualified data should be provided to improve the prediction accuracy on drug-target interactions, which is a key step for drug discovery. 
The number of experimental data samples in the PDB-bind database \cite{liu2017forging} is only about 15,000, limiting the development of reliable deep learning models.
In order to secure more qualified data, expensive and time-consuming experiments are inevitable.
Synthetic data can be used as an alternative like in the DUD-E dataset \cite{mysinger2012directory}, but it may contain unintentional errors.
In addition, data bias and noise often hurt data quality.
The Tox21 dataset \cite{mayr2016deeptox} is such an example.
The number of data samples in Tox21 data set is less than 10,000. 
There are far more negative samples than positive samples. 
Of various toxic types, the lowest percentage of positive samples is 2.9\% and the highest is 15.5\%.
All of those situations would provoke uncertainties of prediction. 
Therefore, uncertainty analysis should be considered to assess deep learning models which must rely on insufficient data and incomplete models. 

In this paper, we propose to use Bayesian neural networks (BNNs) to analyze uncertainties implied in molecular property predictions. 
Quantitative uncertainty analysis enables us to separate model- and data-driven uncertainties, which helps evaluate the reliability of prediction results. 
It is possible because Bayesian inferences allow uncertainty assessments, giving probabilistic interpretations of model outputs. 
Previous studies on uncertainty quantification have utilized the variance of predictive distribution as predictive uncertainty \cite{kendall2017uncertainties, kwon2018uncertainty}. 
The predictive uncertainty can be decomposed into i) an aleatoric uncertainty arisen from data noise and ii) an epistemic uncertainty arisen from the incompleteness of model \cite{der2009aleatory}. 
Thus, we adopted the same method in this study. As a deep learning model for molecular applications, we used graph convolutional networks (GCNs) \cite{, duvenaud2015convolutional, kipf2016semi}. 

In what follows, we will briefly introduce about BNNs, uncertainty quantification methods via Bayesian predictive inferences, and the GCN used in this work. 
Then, we will show the results of uncertainty analysis on three example studies using Bayesian GCN as summarized below.
\begin{itemize}
    \item We first applied the Bayesian GCN to a simple example, logP prediction of molecules in the ZINC set \cite{irwin2005zinc}, in order to validate the uncertainty quantification in molecular applications. As expected, the aleatoric uncertainty increases as the data noise increases, while the epistemic uncertainty does not depend on the quality of data.  
    \item Second, we show that the uncertainty quantification can be used to evaluate the quality of sythetic data. The Harvard Clean Energy Project (CEP) data set \cite{hachmann2011harvard} contains synthetic power conversion efficiency (PCE) values of molecules. We noted that molecules with exactly zero values have conspicuously large aleatoric uncertainty, making us suspect incorrect annotations.
    \item Finally, we studied the relation between predicted values and uncertainties in the binary classifications of bio-activity and toxicity. Our analysis shows that prediction with a lower uncertainty turned out to be more accurate, indicating that the uncertainty can be regarded as the confidence of prediction.  
\end{itemize}

\section{Theoretical backgrounds}
\subsection{Bayesian neural network}
Under the Bayesian framework, model parameters and outputs are considered as random variables. To be specific, we start from defining some notations. For a given training set $\{ \textbf{X}, \textbf{Y} \}$, let  $p(\textbf{Y}|\textbf{X},\textbf{w})$ and $ p(\textbf{w})$ be a BNN model likelihood and a prior for a trainable parameter $\textbf{w} \in \Omega$, respectively. Then, the posterior is given by
\begin{equation} \label{eq:1}
p(\textbf{w}|\textbf{X},\textbf{Y}) = \frac{p(\textbf{Y}|\textbf{X},\textbf{w})p(\textbf{w})}{p(\textbf{Y}|\textbf{X})}
\end{equation}
and the predictive distribution is defined as
\begin{equation} \label{eq:2}
p(\textbf{y}^*|\textbf{x}^*,\textbf{X},\textbf{Y}) = \int _\Omega p(\textbf{y}^*|\textbf{x}^*,\textbf{w}) p(\textbf{w}|\textbf{X},\textbf{Y}) d\textbf{w}
\end{equation}
for a new input $\textbf{x}^*$ and an output $\textbf{y}^*$. These simple formulations show the two fundamental Bayesian inferences: (i) assessing uncertainty of the random variables and (ii) predicting a new output $\textbf{y}^*$ given both a new input $\textbf{x}^*$ and the training set $\{ \textbf{X}, \textbf{Y} \}$.

However, direct exploitation of the formulations is not practical because the integration over the whole parameter space $\Omega$ entails heavy computational costs.
Practical approximations have been studied to resolve the computational limitations. A variational inference, one of the most popular approximation methods, approximates the posterior with a tractable distribution $q_{\theta}(\textbf{w})$ parametrized by a variational parameter $\theta$ \cite{blundell2015weight, graves2011practical}.
Minimizing the Kullback-Leibler divergence, 
\begin{equation} \label{eq:3}
\textrm{KL}(q_{\theta}(\textbf{w}) \Vert p(\textbf{w}|\textbf{X},\textbf{Y})) = \int _\Omega q_{\theta}(\textbf{w}) \log \frac{q_{\theta}(\textbf{w})}{p(\textbf{w}|\textbf{X},\textbf{Y})} d\textbf{w},
\end{equation}
makes the two distributions similar to one another in principle. We can replace the intractable posterior distribution in \eqref{eq:3} with $p(\textbf{Y}|\textbf{X},\textbf{w})p(\textbf{w})$ due to the Bayes' theorem \eqref{eq:1}. Then, our minimization objective with the variational approximation, which is the negative evidence lower-bound, becomes 
\begin{equation} \label{eq:4}
\mathcal{L}_{\textrm{VI}}(\theta) = -\int _\Omega q_{\theta}(\textbf{w})\log{p(\textbf{Y}|\textbf{X},\textbf{w})}d\textbf{w}+\textrm{KL}(q_{\theta}(\textbf{w}) \Vert p(\textbf{w})).
\end{equation}

In order to implement practical Bayesian models, we need to be cautious in choosing a class of variational distribution $q_{\theta}(\textbf{w})$. 
A product of Gaussian distributions was used to approximate the posterior in \cite{blundell2015weight}. However, using fully factorized Gaussian distributions works badly in practice because a posterior can be often a multi-modal distribution. 
A multiplicative normalizing flow \cite{louizos2017multiplicative} can be used for more expressive power, however, this approach requires a large number of weight parameters. 
A dropout network \cite{gal2016dropout} approximates the posterior with using dropout \cite{srivastava2014dropout} in a given neural network. 
The dropout network is practical, because it does not need extra learnable parameters to model the variational posterior distribution and the integration over whole parameter space can be easily approximated with summation of models sampled by a Monte Carlo (MC) estimator. 
See \cite{gal2016uncertainty, gal2016dropout} for more details. Thus, we adopted the dropout network in this work. 

\subsection{Uncertainty in the Bayesian neural network}
A variational inference approximating posterior with  variational distributions $q_{\theta}(\textbf{w})$ provides a variational predictive distribution of a new output $\textbf{y}^*$ given a new input $\textbf{x}^*$ as 
\begin{equation} \label{eq:5}
q_{\theta}^*(\textbf{y}^*|\textbf{x}^*) = \int _\Omega q_{\theta}(\textbf{w}) p(\textbf{y}^*|f^{\textbf{w}}(\textbf{x}^*)) d\textbf{w},
\end{equation}
where $f^{\textbf{w}}(\textbf{x}^*)$ is an output from the model with a given $\textbf{w}$. For regression tasks, a predictive mean of this distribution with $T$ times of MC sampling is estimated by
\begin{equation} \label{eq:6}
    \hat{E} [\textbf{y}^*|\textbf{x}^*] = \frac{1}{T}\sum_{i=1}^{T} f^{\hat{\textbf{w}}_t}(\textbf{x}^*),
\end{equation}
and a predictive variance is estimated by
\begin{equation} \label{eq:7}
    \widehat{Var} [\textbf{y}^*|\textbf{x}^*] = \sigma^2I + \frac{1}{T} \sum_{t=1}^{T} f^{\hat{\textbf{w}}_t}(\textbf{x}^*)^T f^{\hat{\textbf{w}}_t}(\textbf{x}^*) - \hat{E} [\textbf{y}^* | \textbf{x}^*]^T  \hat{E} [\textbf{y}^* | \textbf{x}^*],
\end{equation}
with $\hat{\textbf{w}}_t$ drawn from $q_{\theta}(\textbf{w})$ at the sampling step $t$ and an assumption $p(\textbf{y}^*|f^{\textbf{w}}(\textbf{x}^*)) = N(\textbf{y}^*;f^{\textbf{w}}(\textbf{x}^*), \sigma^2I)$. Here, the model assumes a homoscedasticity with a known quantity, meaning that every data point gives a distribution with a same variance $\sigma$.
Further to this, obtaining the distributions with different variances makes us possible to deduce a heteroscedastic uncertainty. Assuming the heteroscedasticity, the output given the $t$-th sample $\hat{\textbf{w}}_t$ is 
\begin{equation} \label{eq:8}
    [\hat{\textbf{y}}_{t}^*, \hat{\sigma}_{t}] = f^{\hat{\textbf{w}}_t}(\textbf{x}^*).
\end{equation}
The heteroscedastic predictive uncertainty is given by \eqref{eq:9} that can be partitioned into two different uncertainties: aleatoric and epistemic uncertainties.
\begin{equation} \label{eq:9} 
    \widehat{Var} [\textbf{y}^*|\textbf{x}^*] = \underbrace{\frac{1}{T} \sum_{t=1}^{T} (\hat{\textbf{y}}_{t}^{*})^2 - (\frac{1}{T} \sum_{t=1}^{T} \hat{\textbf{y}}_{t}^*)^2}_\textrm{epistemic} + \underbrace{\frac{1}{T} \sum_{t=1}^{T} \hat{\sigma}_{t}^2}_\textrm{aleatoric}.
\end{equation}
The aleatoric uncertainty arises from data inherent noise, while the epistemic uncertainty is related to the model incompleteness. Note that the latter can be reduced by increasing the amount of training data, because it comes from insufficient amount of data as well as the use of inappropriate model \cite{der2009aleatory}.

Furthermore, in classification problems, the authors of \cite{kwon2018uncertainty} proposed a natural way to quantify uncertainties as follows.
\begin{equation} \label{eq:10} 
\widehat{Var} [\textbf{y}^* | \textbf{x}^*] = \underbrace{\frac{1}{T} \sum_{t=1}^{T}(\hat{\textbf{y}}_t^* - \bar{\textbf{y}})(\hat{\textbf{y}}_t^* - \bar{\textbf{y}})^T}_\textrm{epistemic} + \underbrace{\frac{1}{T} \sum_{t=1}^{T} (\textrm{diag}(\hat{\textbf{y}}_t^*) - (\hat{\textbf{y}}_t^*)(\hat{\textbf{y}}_t^*)^T)}_\textrm{aleatoric}, 
\end{equation}
where $\bar{\textbf{y}} = \sum_{t=1}^{T} \hat{\textbf{y}}_t^*/T$ and $\hat{\textbf{y}}_t^* = \textrm{softmax}(\textbf{f}^{\hat{\textbf{w}}_t}(\textbf{x}^*))$. While \cite{kendall2017uncertainties} requires extra parameters $\hat{{\sigma}}_t$ at the last hidden layer and often causes unstable training, \cite{kwon2018uncertainty} has advantages in that models do not involve the extra parameters. The equation \eqref{eq:10} also models functional relationship between mean and variance of multinomial random variables. See \cite{kwon2018uncertainty} for more details.

\subsection{Graph convolutional network for molecular property predictions}

Molecules, social graphs, images and language sentences can be represented as graph structures \cite{battaglia2018relational}. GCN is one of the most popular graph neural networks. Inputs to the GCN is $\textbf{G}=(\textbf{A},\textbf{X})$, where $\textbf{A}$ is an adjacency matrix and $\textbf{X} = \textbf{H}^{(0)}$ is initial node features with the number of nodes $N$. 
The GCN turns out new node features as follows.
\begin{equation} \label{eq:11}
\textbf{H}^{(l+1)} = \textrm{ReLU}(\textbf{A}\textbf{H}^{(l)}\textbf{W}^{(l)}),
\end{equation}
where $\textbf{H}^{(l)}$ and $\textbf{W}^{(l)}$ are node features and trainable parameters for the $l$-th graph convolution layer for $l \in \{ 0, \dots, L\}$, respectively.
The GCN updates node features $\textbf{H}^{(l+1)}$ with information only of adjacent nodes.
Applying attention mechanism enables the GCN to learn relations between node pairs by reflecting the importance of adjacent nodes \cite{velickovic2017graph}. Updating node features with the attention mechanism is given by
\begin{equation} \label{eq:12}
\textbf{H}_i^{(l+1)} = \textrm{ReLU}(\sum_{j \in \mathcal{N}_i}\alpha_{ij} \textbf{H}_j^{(l)} \textbf{W}^{(l)}),  \quad   \alpha_{ij} = \textrm{tanh}((\textbf{H}_i^{(l)}\textbf{W}^{(l)}) \textbf{C}^{(l)}  (\textbf{H}_j^{(l)}\textbf{W}^{(l)})^T),
\end{equation}
where $\mathcal{N}_i$ and $\textbf{C}^{(l)}$ denote the neighbors of the $i$-th node and a trainable parameter, respectively.
In addition, the GCN has a room for an improvement because its accuracy gradually lowered as the number of graph convolution layers increases. \cite{kipf2016semi, ryu2018deeply}. We used a gated-skip connection to prevent this problem as follows.
\begin{equation} \label{eq:13}
\textbf{H}^{(l+1)}_{gsc} = \textbf{z} \odot \textbf{H}^{(l+1)} + (\textbf{1}-\textbf{z}) \odot \textbf{H}^{(l)}, \quad \textbf{z} = \textrm{sigmoid}(\textbf{U}_{z,1} \textbf{H}^{(l)} + \textbf{U}_{z,2}\textbf{H}^{(l+1)} + \textbf{b}_z),
\end{equation}
where $\odot$ denotes Hadamard product. 
After updating the node features $L$-times through feed forward computations, a graph feature $\textbf{z}_{\textrm{G}}$ is summarized as the summation of all node features in a set of nodes $\mathcal{N}$, 
\begin{equation} \label{eq:14}
\textbf{z}_{\textrm{G}} = \sum_{n \in \mathcal{N}} \textrm{NN}(\textbf{H}_n^{(L)}).   
\end{equation}
The graph feature is invariant to permutations of the node states. A molecular property, which is the final output from the neural network, is a function of the graph feature.
\begin{equation} \label{eq:15}
y_{pred} = \textrm{NN}(\textbf{z}_{\textrm{G}}),
\end{equation}
where $\textrm{NN}$ denotes a neural network.


\section{Implementation details}
\subsection{Model architecture}


\begin{figure}
    \centering
    \includegraphics[width=0.9\textwidth]{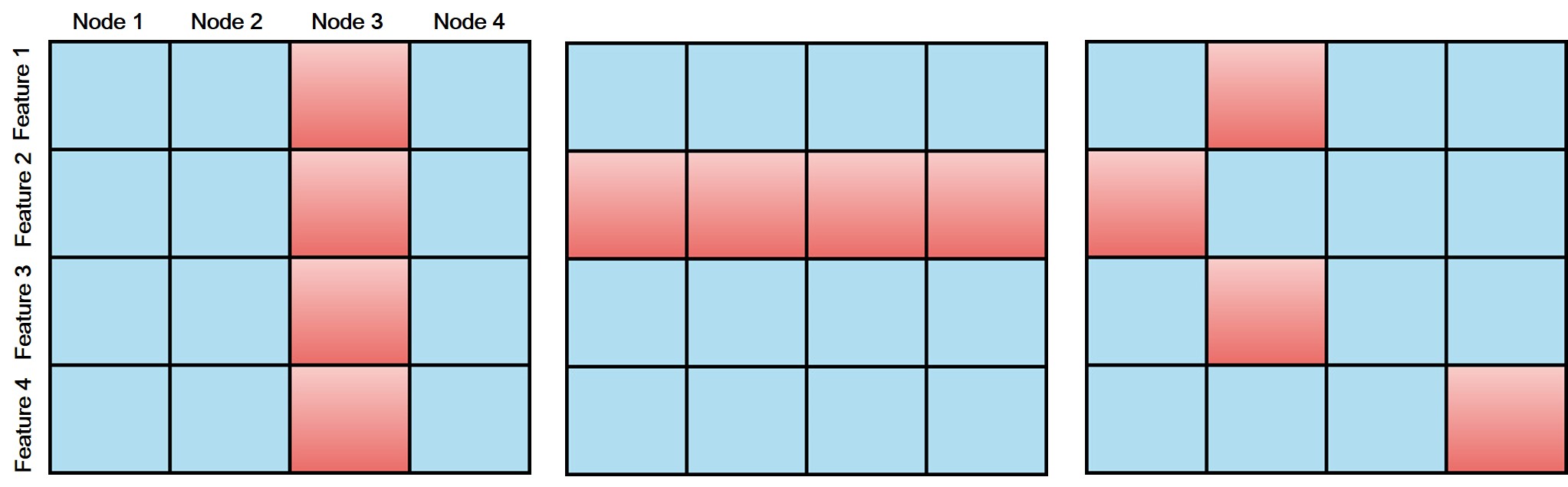}
    \caption{Masking node vectors (\textit{left}), feature vectors (\textit{middle}) and randomly (\textit{right}). We color the masked elements in red. In this illustration, the dropout rate is 0.25.}
    \label{fig:1}
\end{figure}

Our Bayesian neural network used in this work consists of the following three parts:
\begin{itemize}
    \item Three graph convolution layers update node features according to \eqref{eq:13}. The dimension of output from each layer is (number of atoms $\times$ number of features) = ($50\times64$). 
    \item A readout function, which is a single fully-connected layer with a sigmoid activation, produces a graph feature whose dimension is 512. 
    \item A feed-forward neural network composed of three fully-connected layers turns out a molecular property. The hidden dimension of each fully-connected layer is 512. In classification tasks, we used a sigmoid activation, instead of a softmax activation, in the last hidden layer.
\end{itemize}
In order for the model parameters to be stochastic, we applied dropouts at every hidden layer. 
When applying the dropouts, the weight distributions will change according to the dimension of dropout mask. 
We depict the masked shapes of node features. 
We implemented masking the node features as depicted in Figure \ref{fig:1} \textit{middle}.


\subsection{Training conditions}
We manually grid-searched to find the best dropout rates and regularization coefficient for each task. For all experiments, we used Adam optimizer \cite{kingma2014adam} with an initial learning rate $10^{-3}$ and an exponential decay rate $0.95$. The number of epoch was $150$ and the batch size was $100$.

\section{Experiments}

\subsection{Aleatoric uncertainty due to data quality}
 
\begin{figure}
    \centering
    \includegraphics[width=1.0\textwidth]{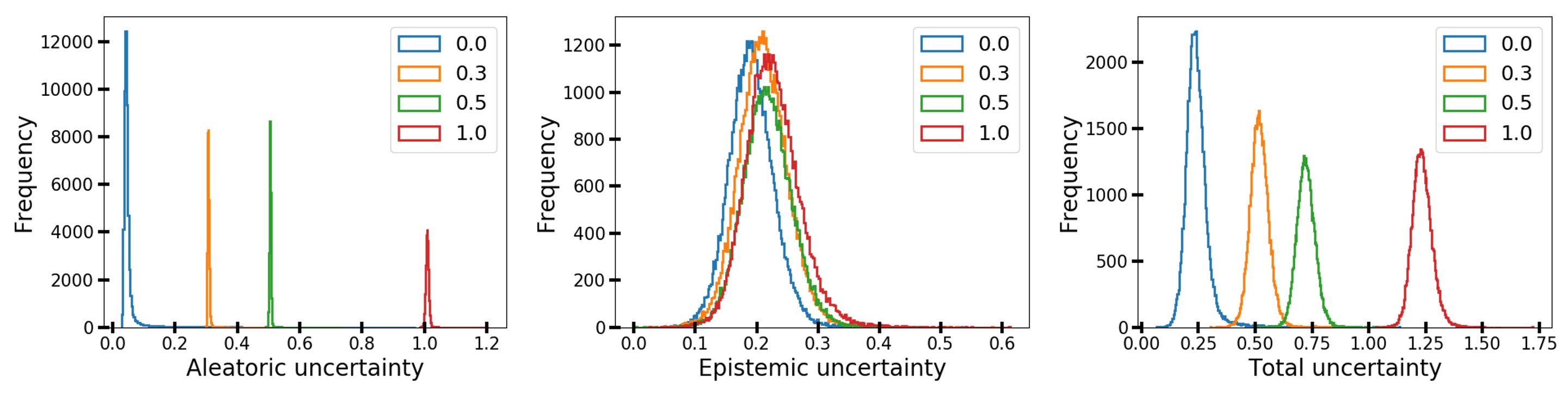}
    \caption{Histograms of aleatoric (\textit{left}), epistemic (\textit{middle}) and total (\textit{right}) uncertanties as the amount of additive noise $\sigma^2$ increases. }
    \label{fig:4}
\end{figure}

In this experiment, we applied the uncertainty quantification method to a simple example, logP prediction.
We chose this example because we can obtain the logP values of molecules from analytic expression of logP as implemented in the RDKit \cite{landrum2006rdkit} without data inherent noise. 
To examine the effect of data quality on uncertainties, we could adjust the extent of noise in logP by adding a random Gaussian noise $\epsilon \sim \mathcal{N}(0,\sigma^2)$. 
We trained the model with 240k samples and analyzed uncertainties of each predicted logP for 50k samples. The samples were chosen randomly from the ZINC set \textcolor{red}.

Figure \ref{fig:4} shows the distribution of the three uncertainties as a function of the amount of additive noise $\sigma^2$. 
As the noise level increases, the aleatoric and total uncertainties increase, but the epistemic uncertainty is almost unchanged. This result confirms that the aleatoric uncertainty indeed arises due to data inherent noises, while the epistemic uncertainty does not depend on data quality. 

\subsection{Evaluating quality of synthetic data based on uncertainty analysis}


\begin{figure}
    \centering
    \includegraphics[width=0.8\textwidth]{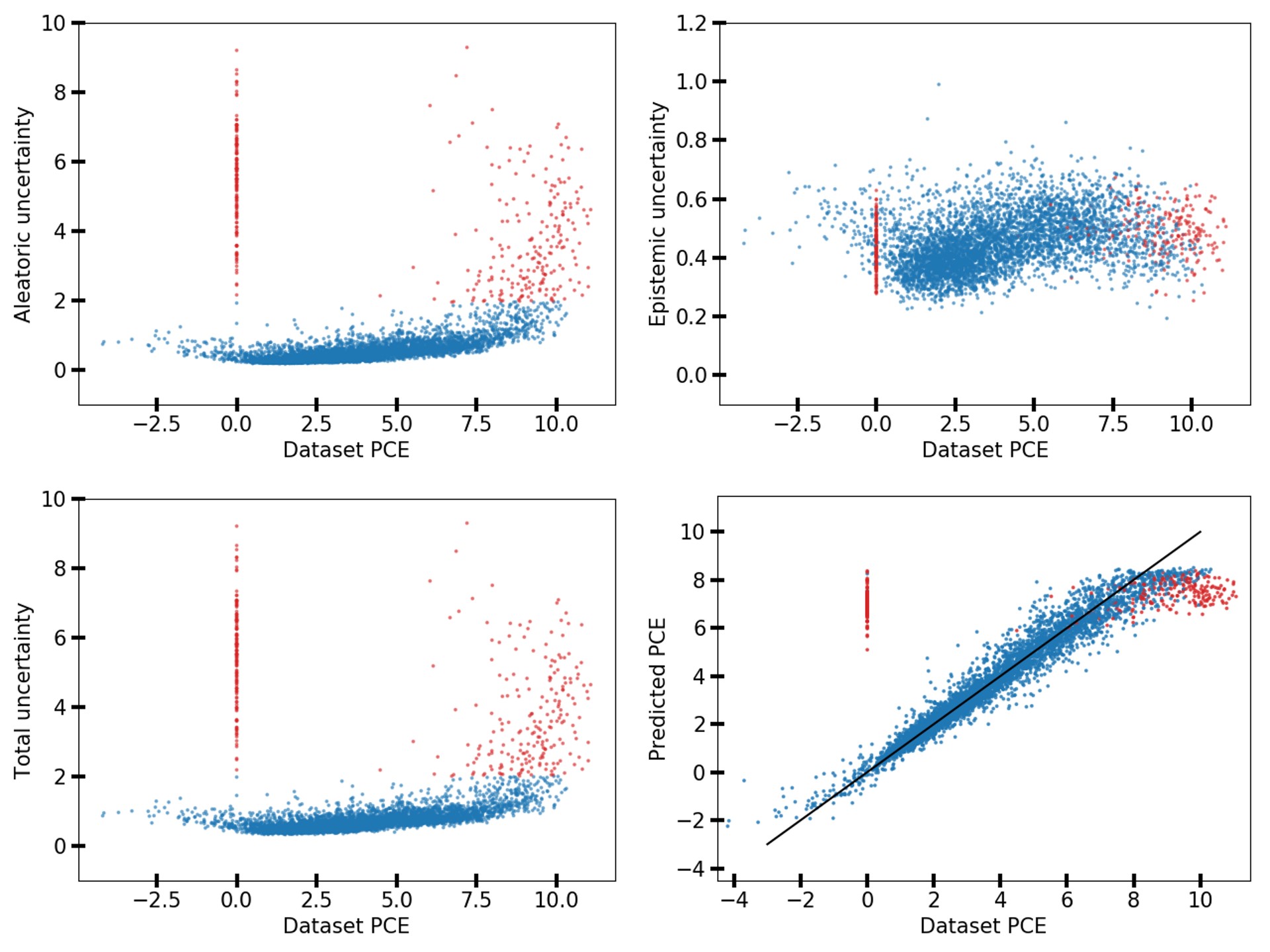}
    \caption{Aleatoric (\textit{top, left}), epistemic (\textit{top, right}), total uncertainties (\textit{bottom, left}) and predicted PCE (\textit{bottom, right}) against the dataset PCE. The samples colored in red show the total uncertainty greater than two.}
    \label{fig:5}
\end{figure}

Synthetic PCE values in the CEP dataset \cite{hachmann2011harvard} was obtained from the Scharber model with statistical approximations \cite{scharber2006design}. In this procedure, unintended errors can be included in the resulting synthetic data. Since the aleatoric uncertainty is due to data quality, we evaluated quality of the synthetic data by analyzing the uncertainties of predicted PCE values. We used the same dataset in \cite{duvenaud2015convolutional}\footnote{https://github.com/HIPS/neural-fingerprint} for training and test.

Figure \ref{fig:5} shows the distribution of the three uncertainties of each molecule in the test set.
Samples with the total uncertainty greater than two were highlighted with red color. 
Some samples with large PCE values above eight had relatively large total uncertainties. Their PCE values deviated considerably from the black line in Figure \ref{fig:4} (\textit{bottom}, \textit{right}). More interestingly, we noted that most molecules with the zero PCE value had large total uncertainties as well. Those large uncertainties came from the aleatoric uncertainty as depiced in in Figure \ref{fig:4} (\textit{top}, \textit{left}), indicating that the data quality of those particular samples is relatively bad. Hence, we speculated that the inaccurate predictions are due to data inherent noise. 

To elaborate the origin of such large errors, we investigated the procedure of obtaining the PCE values. The Havard Organic Photovolatic Dataset \cite{lopez2016harvard} contains both experimental and synthetic PCE values of 350 organic photovoltaic materials.
The synthetic PCE values were computed according to \eqref{eq:16}, which is the result of the Scharber model \cite{scharber2006design}. 
\begin{equation}\label{eq:16}
\textrm{PCE} \propto V_{OC} \times FF \times J_{SC},
\end{equation}
where $V_{OC}$ is an open circuit potential, $FF$ is a fill factor, and $J_{SC}$ is short circuit current density.
$FF$ was set to 65\%. $V_{OC}$ and $J_{SC}$ were obtained from electronic structure calculations of molecules. See \cite{hachmann2011harvard} for more details.
We found that $J_{SC}$ of some molecules are zero or nearly zero, resulting in zero or almost zero synthetic PCE values, in contrast to their non-zero experimental PCE values.
Especially, $J_{SC}$ and PCE values computed using the M06-2X functional \cite{zhao2008m06} were almost zero consistently.
We suspect that those approximated values caused a significant drop of data quality, resulting in large aleatoric uncertainties as highlighted in Figure \ref{fig:4}. Consequently, data noise due to badly fabricated data was identified as the large aleatoric uncertainties.

\subsection{Uncertainty as confidence indicator: bio-activity and toxicity classifications}

Many molecular property datasets are provided with binary or multi-class labels. Accurate classification is vital for medical applications of molecules, such as bio-activity and toxicity for drug discovery. In this experiment, we assessed whether or not the uncertainty analysis can be used to help more accurate classifications.  

\begin{figure}
    \centering
    \includegraphics[width=1.0\textwidth]{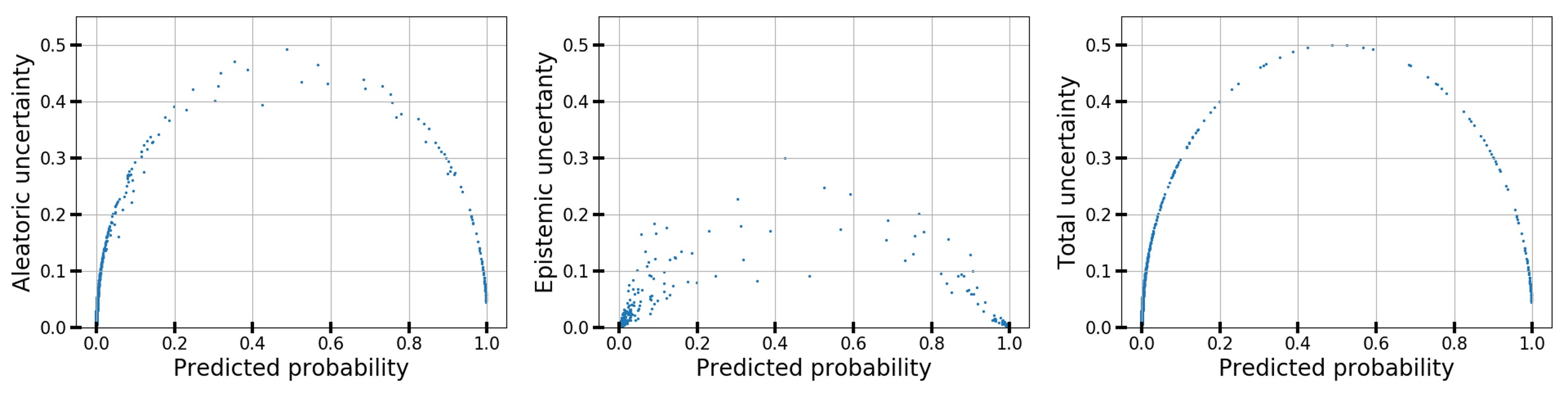}
    \caption{Aleatoric (\textit{left}), epistemic (\textit{middle}) and  total uncertainty (\textit{right}) of predicted labels for the classification of bio-activity to the EGFR target.}
    \label{fig:6}
\end{figure}

In classifications, sigmoid activated outputs can be interpreted as confidence if outputs near zero/one are closer to their ground truth label zero/one and those near 0.5 are uncertain to determine their truth labels. 
In this context, it is expected that outputs near zero/one ought to have lower uncertainties, whereas those near 0.5 ought to have higher uncertainties. 
We examined if the size of predicted labels in a binary classification problem is related to confidence in correct labeling.
Figure \ref{fig:6} shows the results of bio-activity classification of DUD-E dataset molecules against the EGFR target protein.
Indeed, the uncertainty was minimum on the predicted labels of zero or one and gradually reached the maximum value on the middle.   
We also noted that the aleatoric uncertainty affected the total uncertainty more significantly than the epistemic uncertainty did.

\begin{figure}
    \centering
    \includegraphics[width=1.0\textwidth]{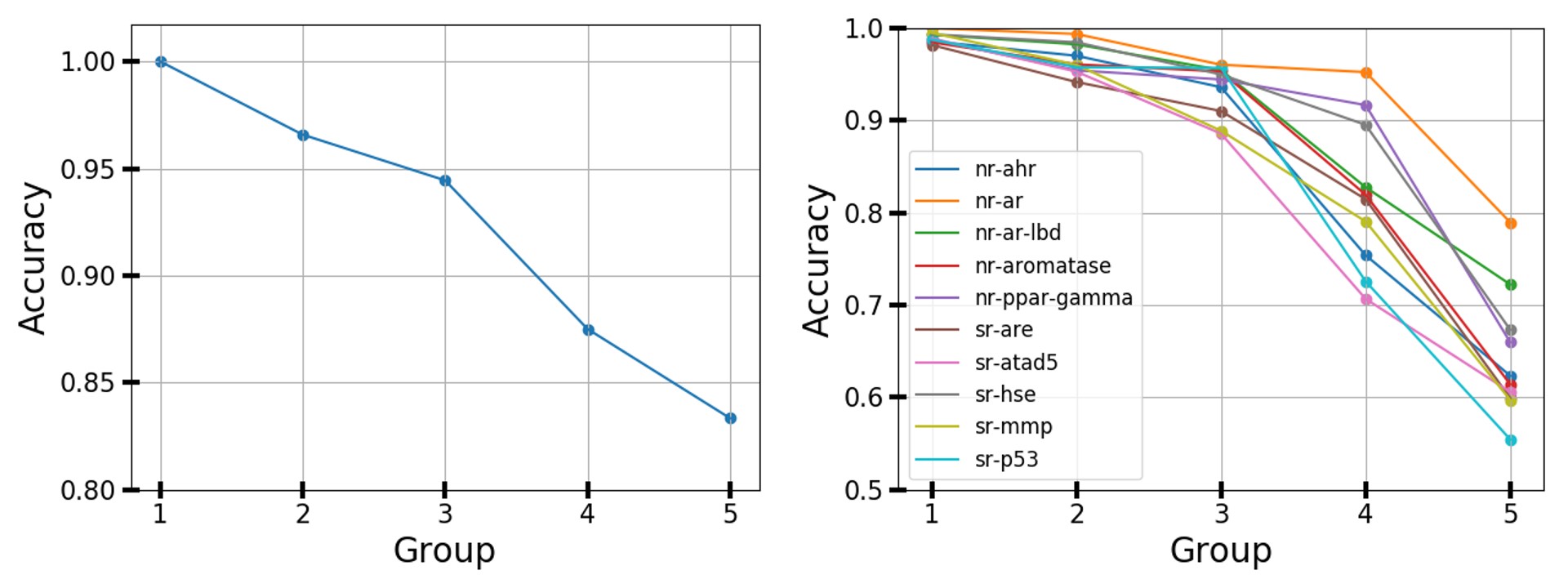}
    \caption{Test accuracy for the classifications of bio-activity to the EGFR target (\textit{left}) and various toxic effects. (\textit{right}).}
    \label{fig:7}
\end{figure}

To further investigate a relation between accuracy and uncertainty, we sorted the molecules in the order of increasing uncertainty and then divided them into five groups.
Figure \ref{fig:7} shows the classification accuracy of each group; the \textit{left} and \textit{right} figures denote the classificaiton results of bio-acitvity and ten different toxicities of Tox21 dataset molecules, respectively. 
Groups with lower uncertainties had higher accuracy.
This result is an evidence that the uncertainty can be used as a confidence indicator in binary classification problems.

In this work, our predicted binary labels come out from the sigmoid activation and labels with 0.5 have the highest total uncertainty. On the other hand, a softmax activation is used in multinomial classifications. 
Softmax activated outputs tend to be interpreted as confidence, which means that the higher the output probability, the higher the prediction accuracy.
However, \cite{gal2016dropout} showed that a larger output value can have a higher uncertainty. 

\section{Conclusion and discussion}

In this work, we analyzed uncertainties in prediction of molecular properties using a Bayesian graph convolutional network. Our first experiment on the logP prediction showed that the aleatoric uncertainty can be used to quantify the extent of data noise. In the second experiment, we could identify badly approximated synthetic data in the Harvard Clean Energy Project dataset by analyzing the aleatoric uncertainty. The last example shows that the uncertainty is closely related to the confidence of prediction in a binary classification problem. All the results demonstrate how useful the uncertainty quantification is in molecular applications as a data quality checker and a confidence indicator.

The Bayesian neural network can be further improved in a technical aspect toward more reliable molecular applications. 
We determined dropout rates and regularization coefficients by manual grid searching. The resulting dropout rates may not be optimal. 
Optimal dropout rates would depend on model architectures and might be different for each hidden layer in a given model. 
For example, a model with a small number of parameters may need small dropout rates.
One can determine optimal dropout rates of each hidden layer in a given model architecture by adopting `Concrete dropout' \cite{gal2017concrete}. 
It has been shown that a Bayesian model with the Concrete dropout outperforms a model with manually-tuned dropout rates. 
Thus, we can implement a more sophisticated Bayesian model by utilizing the Concrete dropout. 
In the near future, we will further study the uncertainty quantification of molecular property predictions in order to achieve precise molecular applications.

\subsubsection*{Acknowledgments}

This work was supported by the National Research Foundation of Korea (NRF) grant funded by the Korea government (MSIT)(NRF-2017R1E1A1A01078109)

\medskip
\small


\begin{thebibliography}{10}
	
	\bibitem{battaglia2018relational}
	Peter~W Battaglia, Jessica~B Hamrick, Victor Bapst, Alvaro Sanchez-Gonzalez,
	Vinicius Zambaldi, Mateusz Malinowski, Andrea Tacchetti, David Raposo, Adam
	Santoro, Ryan Faulkner, et~al.
	\newblock Relational inductive biases, deep learning, and graph networks.
	\newblock {\em arXiv preprint arXiv:1806.01261}, 2018.
	
	\bibitem{blundell2015weight}
	Charles Blundell, Julien Cornebise, Koray Kavukcuoglu, and Daan Wierstra.
	\newblock Weight uncertainty in neural networks.
	\newblock {\em arXiv preprint arXiv:1505.05424}, 2015.
	
	\bibitem{de2018molgan}
	Nicola De~Cao and Thomas Kipf.
	\newblock Molgan: An implicit generative model for small molecular graphs.
	\newblock {\em arXiv preprint arXiv:1805.11973}, 2018.
	
	\bibitem{der2009aleatory}
	Armen Der~Kiureghian and Ove Ditlevsen.
	\newblock Aleatory or epistemic? does it matter?
	\newblock {\em Structural Safety}, 31(2):105--112, 2009.
	
	\bibitem{duvenaud2015convolutional}
	David~K Duvenaud, Dougal Maclaurin, Jorge Iparraguirre, Rafael Bombarell,
	Timothy Hirzel, Al{\'a}n Aspuru-Guzik, and Ryan~P Adams.
	\newblock Convolutional networks on graphs for learning molecular fingerprints.
	\newblock In {\em Advances in neural information processing systems}, pages
	2224--2232, 2015.
	
	\bibitem{faber2017prediction}
	Felix~A Faber, Luke Hutchison, Bing Huang, Justin Gilmer, Samuel~S Schoenholz,
	George~E Dahl, Oriol Vinyals, Steven Kearnes, Patrick~F Riley, and O~Anatole
	von Lilienfeld.
	\newblock Prediction errors of molecular machine learning models lower than
	hybrid dft error.
	\newblock {\em Journal of chemical theory and computation}, 13(11):5255--5264,
	2017.
	
	\bibitem{feinberg2018spatial}
	Evan~N Feinberg, Debnil Sur, Brooke~E Husic, Doris Mai, Yang Li, Jianyi Yang,
	Bharath Ramsundar, and Vijay~S Pande.
	\newblock Spatial graph convolutions for drug discovery.
	\newblock {\em arXiv preprint arXiv:1803.04465}, 2018.
	
	\bibitem{gal2016uncertainty}
	Yarin Gal.
	\newblock Uncertainty in deep learning.
	\newblock {\em University of Cambridge}, 2016.
	
	\bibitem{gal2016dropout}
	Yarin Gal and Zoubin Ghahramani.
	\newblock Dropout as a bayesian approximation: Representing model uncertainty
	in deep learning.
	\newblock In {\em international conference on machine learning}, pages
	1050--1059, 2016.
	
	\bibitem{gal2017concrete}
	Yarin Gal, Jiri Hron, and Alex Kendall.
	\newblock Concrete dropout.
	\newblock In {\em Advances in Neural Information Processing Systems}, pages
	3581--3590, 2017.
	
	\bibitem{gilmer2017neural}
	Justin Gilmer, Samuel~S Schoenholz, Patrick~F Riley, Oriol Vinyals, and
	George~E Dahl.
	\newblock Neural message passing for quantum chemistry.
	\newblock {\em arXiv preprint arXiv:1704.01212}, 2017.
	
	\bibitem{gomes2017atomic}
	Joseph Gomes, Bharath Ramsundar, Evan~N Feinberg, and Vijay~S Pande.
	\newblock Atomic convolutional networks for predicting protein-ligand binding
	affinity.
	\newblock {\em arXiv preprint arXiv:1703.10603}, 2017.
	
	\bibitem{gomez2018automatic}
	Rafael G{\'o}mez-Bombarelli, Jennifer~N Wei, David Duvenaud, Jos{\'e}~Miguel
	Hern{\'a}ndez-Lobato, Benjam{\'\i}n S{\'a}nchez-Lengeling, Dennis Sheberla,
	Jorge Aguilera-Iparraguirre, Timothy~D Hirzel, Ryan~P Adams, and Al{\'a}n
	Aspuru-Guzik.
	\newblock Automatic chemical design using a data-driven continuous
	representation of molecules.
	\newblock {\em ACS central science}, 4(2):268--276, 2018.
	
	\bibitem{graves2011practical}
	Alex Graves.
	\newblock Practical variational inference for neural networks.
	\newblock In {\em Advances in neural information processing systems}, pages
	2348--2356, 2011.
	
	\bibitem{guimaraes2017objective}
	Gabriel~Lima Guimaraes, Benjamin Sanchez-Lengeling, Carlos Outeiral, Pedro
	Luis~Cunha Farias, and Al{\'a}n Aspuru-Guzik.
	\newblock Objective-reinforced generative adversarial networks (organ) for
	sequence generation models.
	\newblock {\em arXiv preprint arXiv:1705.10843}, 2017.
	
	\bibitem{hachmann2011harvard}
	Johannes Hachmann, Roberto Olivares-Amaya, Sule Atahan-Evrenk, Carlos
	Amador-Bedolla, Roel~S S{\'a}nchez-Carrera, Aryeh Gold-Parker, Leslie Vogt,
	Anna~M Brockway, and Al{\'a}n Aspuru-Guzik.
	\newblock The harvard clean energy project: large-scale computational screening
	and design of organic photovoltaics on the world community grid.
	\newblock {\em The Journal of Physical Chemistry Letters}, 2(17):2241--2251,
	2011.
	
	\bibitem{irwin2005zinc}
	John~J Irwin and Brian~K Shoichet.
	\newblock Zinc- a free database of commercially available compounds for virtual
	screening.
	\newblock {\em Journal of chemical information and modeling}, 45(1):177--182,
	2005.
	
	\bibitem{jimenez2018k}
	Jos{\'e} Jim{\'e}nez, Miha Skalic, Gerard Mart{\'\i}nez-Rosell, and Gianni
	De~Fabritiis.
	\newblock K deep: Protein--ligand absolute binding affinity prediction via
	3d-convolutional neural networks.
	\newblock {\em Journal of chemical information and modeling}, 58(2):287--296,
	2018.
	
	\bibitem{jin2018junction}
	Wengong Jin, Regina Barzilay, and Tommi Jaakkola.
	\newblock Junction tree variational autoencoder for molecular graph generation.
	\newblock {\em arXiv preprint arXiv:1802.04364}, 2018.
	
	\bibitem{kendall2017uncertainties}
	Alex Kendall and Yarin Gal.
	\newblock What uncertainties do we need in bayesian deep learning for computer
	vision?
	\newblock In {\em Advances in neural information processing systems}, pages
	5574--5584, 2017.
	
	\bibitem{kingma2014adam}
	Diederik~P Kingma and Jimmy Ba.
	\newblock Adam: A method for stochastic optimization.
	\newblock {\em arXiv preprint arXiv:1412.6980}, 2014.
	
	\bibitem{kipf2016semi}
	Thomas~N Kipf and Max Welling.
	\newblock Semi-supervised classification with graph convolutional networks.
	\newblock {\em arXiv preprint arXiv:1609.02907}, 2016.
	
	\bibitem{kusner2017grammar}
	Matt~J Kusner, Brooks Paige, and Jos{\'e}~Miguel Hern{\'a}ndez-Lobato.
	\newblock Grammar variational autoencoder.
	\newblock {\em arXiv preprint arXiv:1703.01925}, 2017.
	
	\bibitem{kwon2018uncertainty}
	Yongchan Kwon, Joong-Ho Won, Beom~Joon Kim, and Myunghee~Cho Paik.
	\newblock Uncertainty quantification using bayesian neural networks in
	classification: Application to ischemic stroke lesion segmentation.
	\newblock In {\em international conference on medical imaging with deep
		learning}, 2018.
	
	\bibitem{landrum2006rdkit}
	Greg Landrum et~al.
	\newblock Rdkit: Open-source cheminformatics, 2006.
	
	\bibitem{li2018learning}
	Yujia Li, Oriol Vinyals, Chris Dyer, Razvan Pascanu, and Peter Battaglia.
	\newblock Learning deep generative models of graphs.
	\newblock {\em arXiv preprint arXiv:1803.03324}, 2018.
	
	\bibitem{liu2017forging}
	Zhihai Liu, Minyi Su, Li~Han, Jie Liu, Qifan Yang, Yan Li, and Renxiao Wang.
	\newblock Forging the basis for developing protein--ligand interaction scoring
	functions.
	\newblock {\em Accounts of chemical research}, 50(2):302--309, 2017.
	
	\bibitem{lopez2016harvard}
	Steven~A Lopez, Edward~O Pyzer-Knapp, Gregor~N Simm, Trevor Lutzow, Kewei Li,
	Laszlo~R Seress, Johannes Hachmann, and Al{\'a}n Aspuru-Guzik.
	\newblock The harvard organic photovoltaic dataset.
	\newblock {\em Scientific data}, 3:160086, 2016.
	
	\bibitem{louizos2017multiplicative}
	Christos Louizos and Max Welling.
	\newblock Multiplicative normalizing flows for variational bayesian neural
	networks.
	\newblock {\em arXiv preprint arXiv:1703.01961}, 2017.
	
	\bibitem{mayr2016deeptox}
	Andreas Mayr, G{\"u}ntero Klambauer, Thomas Unterthiner, and Sepp Hochreiter.
	\newblock Deeptox: toxicity prediction using deep learning.
	\newblock {\em Frontiers in Environmental Science}, 3:80, 2016.
	
	\bibitem{mysinger2012directory}
	Michael~M Mysinger, Michael Carchia, John~J Irwin, and Brian~K Shoichet.
	\newblock Directory of useful decoys, enhanced (dud-e): better ligands and
	decoys for better benchmarking.
	\newblock {\em Journal of medicinal chemistry}, 55(14):6582--6594, 2012.
	
	\bibitem{ozturk2018deepdta}
	Hakime {\"O}zt{\"u}rk, Arzucan {\"O}zg{\"u}r, and Elif Ozkirimli.
	\newblock Deepdta: deep drug--target binding affinity prediction.
	\newblock {\em Bioinformatics}, 34(17):i821--i829, 2018.
	
	\bibitem{ryu2018deeply}
	Seongok Ryu, Jaechang Lim, and Woo~Youn Kim.
	\newblock Deeply learning molecular structure-property relationships using
	graph attention neural network.
	\newblock {\em arXiv preprint arXiv:1805.10988}, 2018.
	
	\bibitem{scharber2006design}
	Markus~C Scharber, David M{\"u}hlbacher, Markus Koppe, Patrick Denk, Christoph
	Waldauf, Alan~J Heeger, and Christoph~J Brabec.
	\newblock Design rules for donors in bulk-heterojunction solar cells—towards
	10\% energy-conversion efficiency.
	\newblock {\em Advanced materials}, 18(6):789--794, 2006.
	
	\bibitem{schutt2017schnet}
	Kristof Sch{\"u}tt, Pieter-Jan Kindermans, Huziel Enoc~Sauceda Felix, Stefan
	Chmiela, Alexandre Tkatchenko, and Klaus-Robert M{\"u}ller.
	\newblock Schnet: A continuous-filter convolutional neural network for modeling
	quantum interactions.
	\newblock In {\em Advances in Neural Information Processing Systems}, pages
	991--1001, 2017.
	
	\bibitem{schutt2017quantum}
	Kristof~T Sch{\"u}tt, Farhad Arbabzadah, Stefan Chmiela, Klaus~R M{\"u}ller,
	and Alexandre Tkatchenko.
	\newblock Quantum-chemical insights from deep tensor neural networks.
	\newblock {\em Nature communications}, 8:13890, 2017.
	
	\bibitem{segler2017generating}
	Marwin~HS Segler, Thierry Kogej, Christian Tyrchan, and Mark~P Waller.
	\newblock Generating focused molecule libraries for drug discovery with
	recurrent neural networks.
	\newblock {\em ACS central science}, 4(1):120--131, 2017.
	
	\bibitem{segler2018planning}
	Marwin~HS Segler, Mike Preuss, and Mark~P Waller.
	\newblock Planning chemical syntheses with deep neural networks and symbolic
	ai.
	\newblock {\em Nature}, 555(7698):604, 2018.
	
	\bibitem{smith2017ani}
	Justin~S Smith, Olexandr Isayev, and Adrian~E Roitberg.
	\newblock Ani-1: an extensible neural network potential with dft accuracy at
	force field computational cost.
	\newblock {\em Chemical science}, 8(4):3192--3203, 2017.
	
	\bibitem{srivastava2014dropout}
	Nitish Srivastava, Geoffrey Hinton, Alex Krizhevsky, Ilya Sutskever, and Ruslan
	Salakhutdinov.
	\newblock Dropout: a simple way to prevent neural networks from overfitting.
	\newblock {\em The Journal of Machine Learning Research}, 15(1):1929--1958,
	2014.
	
	\bibitem{velickovic2017graph}
	Petar Velickovic, Guillem Cucurull, Arantxa Casanova, Adriana Romero, Pietro
	Lio, and Yoshua Bengio.
	\newblock Graph attention networks.
	\newblock {\em arXiv preprint arXiv:1710.10903}, 2017.
	
	\bibitem{wei2016neural}
	Jennifer~N Wei, David Duvenaud, and Al{\'a}n Aspuru-Guzik.
	\newblock Neural networks for the prediction of organic chemistry reactions.
	\newblock {\em ACS central science}, 2(10):725--732, 2016.
	
	\bibitem{you2018graph}
	Jiaxuan You, Bowen Liu, Rex Ying, Vijay Pande, and Jure Leskovec.
	\newblock Graph convolutional policy network for goal-directed molecular graph
	generation.
	\newblock {\em arXiv preprint arXiv:1806.02473}, 2018.
	
	\bibitem{zhao2008m06}
	Yan Zhao and Donald~G Truhlar.
	\newblock The m06 suite of density functionals for main group thermochemistry,
	thermochemical kinetics, noncovalent interactions, excited states, and
	transition elements: two new functionals and systematic testing of four
	m06-class functionals and 12 other functionals.
	\newblock {\em Theoretical Chemistry Accounts}, 120(1-3):215--241, 2008.
	
	\bibitem{zhou2017optimizing}
	Zhenpeng Zhou, Xiaocheng Li, and Richard~N Zare.
	\newblock Optimizing chemical reactions with deep reinforcement learning.
	\newblock {\em ACS central science}, 3(12):1337--1344, 2017.
	
\end{thebibliography}
\bibliographystyle{plain}

\end{document}